\documentclass[prb,reprint,floatfix,superscriptaddress]{revtex4-2}

\usepackage{amsmath,amsfonts,amssymb,amsthm,braket,graphicx,color,lipsum}
\usepackage{float}
\usepackage[colorlinks=true,citecolor=blue,linkcolor=magenta]{hyperref}
\usepackage{bm}
\usepackage{aligned-overset}
\usepackage[dvipsnames]{xcolor}

\global\long\def\p{\prime}

\global\long\def\ket#1{|#1\rangle}
\global\long\def\bra#1{\langle#1|}
\global\long\def\proj#1#2{|#1\rangle\langle#2|}
\global\long\def\inner#1#2{\langle#1|#2\rangle}
\global\long\def\tr{\mathrm{tr}}

\global\long\def\im{\imath}

\newcommand{\Hsquare}{
  \text{\fboxsep=.0pt\fbox{\rule{0pt}{1ex}\rule{1ex}{0pt}}}
}

\newcommand{\dg} {{\dagger}}
\newcommand{\pd} {{\phantom\dagger}}

\newcommand{\ci}[1] {{c_{#1}^{\pd}}}
\newcommand{\cid}[1] {c_{#1}^\dg}

\newcommand{\bG} {\bm{G}}
\newcommand{\bS} {\bm{\Sigma}}
\newcommand{\bGa} {\bm{\Gamma}}

\newcommand{\bHS}{\bm{\bar{H}}_\qs}

\renewcommand{\Im} {\operatorname{Im}}

\newcommand{\vFv}{\left| v_F \right|}
\newcommand{\vFs}{\left| v_F \right|^2}
\newcommand{\vv}{\left| v \right|}
\newcommand{\vs}{\left| v \right|^2}
\newcommand{\vf}{\left| v \right|^4}
\newcommand{\vls}{\left| \tilde{v} \right|^2}
\newcommand{\vlf}{\left| \tilde{v}  \right|^4}

\newcommand{\ql}{\mathcal{L}}
\newcommand{\qs}{\mathcal{S}}
\newcommand{\qr}{\mathcal{R}}

\newcommand{\qi}{\mathcal{I}}
\newcommand{\qu}{\mathcal{U}}

\newcommand{\qb}{\mathcal{B}}
\newcommand{\qw}{\mathcal{W}}

\newcommand{\ok}{\proj{v_k}{v_k}}
\newcommand{\bk}{\bra{v_k}}
\newcommand{\kk}{\ket{v_k}}
\newcommand{\ik}{\left| v_k \right|^2}
\newcommand{\oF}{\proj{v_F}{v_F}}
\newcommand{\bF}{\bra{v_F}}
\newcommand{\kF}{\ket{v_F}}

\newcommand{\M}{\mathrm{M}}
\newcommand{\nM}{\mathrm{nM}}
\newcommand{\ES}{\mathrm{ES}}
\newcommand{\OD}{\mathrm{TS}}
\newcommand{\WC}{\mathrm{WC}}
\newcommand{\PC}{\lambda}
\newcommand{\LR}{\underset{\mathrm{lr}}}

\newcommand\trick[1]{}

\renewcommand{\[}{\begin{equation}}
\renewcommand{\]}{\end{equation}}

\begin{document}

\title{Dual current anomalies and quantum transport within extended reservoir simulations}

\author{Gabriela W\'{o}jtowicz}
\affiliation{Jagiellonian University, Institute of Theoretical Physics, \L{}ojasiewicza 11, 30-348 Krak\'{o}w, Poland}
\author{Justin E. Elenewski} 
\affiliation{Biophysical and Biomedical Measurement Group, Microsystems and Nanotechnology Division, Physical Measurement Laboratory, National Institute of Standards and Technology, Gaithersburg, MD, USA}
\affiliation{Institute for Research in Electronics and Applied Physics, University of Maryland, College Park, MD, USA}
\author{Marek M. Rams}
\email{marek.rams@uj.edu.pl}
\affiliation{Jagiellonian University, Institute of Theoretical Physics, \L{}ojasiewicza 11, 30-348 Krak\'{o}w, Poland}
\author{Michael Zwolak}
\email{mpz@nist.gov}
\affiliation{Biophysical and Biomedical Measurement Group, Microsystems and Nanotechnology Division, Physical Measurement Laboratory, National Institute of Standards and Technology, Gaithersburg, MD, USA}

\begin{abstract}
Quantum transport simulations are rapidly evolving and now encompass well--controlled tensor network techniques for many--body limits. One powerful approach combines matrix product states with extended reservoirs.  In this method, continuous reservoirs are represented by explicit, discretized counterparts and a chemical potential or temperature drop is maintained by external relaxation. Currents are strongly influenced by relaxation when it is very weak or strong, resulting in a simulation analog of Kramers' turnover for solution--phase chemical reactions. At intermediate relaxation, the intrinsic conductance, that given by the Landauer or Meir--Wingreen expressions, moderates the current. We demonstrate that strong impurity scattering (i.e., a small steady--state current) reveals anomalous transport regimes within this methodology at weak--to--moderate and moderate--to--strong relaxation. The former is due to virtual transitions and the latter to unphysical broadening of the populated density of states. Thus, the turnover analog has {\em five} standard transport regimes, further constraining the parameters that lead to recovery of the intrinsic conductance. In the worst case, the common strategy of choosing a relaxation strength proportional to the reservoir level spacing can prevent convergence to the continuum limit. This advocates a simulation strategy where one utilizes the current versus relaxation turnover profiles to identify simulation parameters that most efficiently reproduce the intrinsic physical behavior.
\end{abstract}

\maketitle
\section{Introduction}
\label{sec:Intro}

The accurate simulation of many--body transport is essential to  understanding nanoscale electronics and quantum dots~\cite{van_der_wiel_electron_2002, weinmann_transport_2010, cha_quantum_2020}, quantum dynamics and control \cite{mabuchi_principles_2005, ke_electrical_2020, petersen_quantum_2010, chu_cold_2002}, spintronic phenomena \cite{zutic_spintronics_2004, pesin_spintronics_2012,linder_superconducting_2015}, and the development of ``atomtronic'' platforms for physical simulation~\cite{chien_bosonic_2012,brantut_conduction_2012,brantut_thermoelectric_2013,chien_interaction-induced_2013, chien_landauer_2014,krinner_observation_2015,krinner_mapping_2016,krinner_two-terminal_2017,hausler_scanning_2017,gruss_energy-resolved_2018}. Recent developments have delivered increasingly rigorous and well--controlled numerical tools to pursue this goal. One approach employs a canonical transformation to a {\em mixed basis}, where energy or momentum modes are paired according to their natural scattering structure, to perform tensor network simulations that are extensive in space and time~\cite{rams_breaking_2020}. This is a substantial advance for matrix product state calculations, which are otherwise limited by the linear growth of entanglement entropy~\cite{cazalilla_time-dependent_2002,zwolak_mixed-state_2004,gobert_real-time_2005,al-hassanieh_adaptive_2006,schneider_conductance_2006,schmitteckert_signal_2006,dias_da_silva_transport_2008,heidrich-meisner_real-time_2009,branschadel_conductance_2010,chien_interaction-induced_2013,gruss_energy-resolved_2018} (though some operate in linear response via the Kubo formula~\cite{bohr_dmrg_2005,bohr_strong_2007}). Alternative strategies have also been presented, including those that introduce  a linear--logarithmic reservoir discretization and reorganize reservoir modes to improve performance~\cite{schwarz_nonequilibrium_2018}. All of these techniques assume closed systems, which only give quasi--steady--state currents when starting from a density or chemical potential imbalance~\cite{zwolak_communication_2018, bushong_approach_2005}.

The mixed--basis approach reflects a natural scattering structure, making it ideal for simulating open systems.  Notably, it can directly target steady and Floquet states, or simulate real--time noise around a stationary state~\cite{wojtowicz_open-system_2020}.  Related methods have also been applied to describe quantum thermal machines~\cite{brenes_tensor-network_2020}. These developments employ ``extended'' or  ``mesoscopic'' reservoirs, where finite collections of fermionic modes are broadened by external environments to yield an effective continuum. 

The extended reservoir approach (ERA) has a lengthy history, originating in  the relaxation time approximation~\cite{kohn_quantum_1957,frensley_simulation_1985,frensley_boundary_1990,mizuta_transient_1991,fischetti_theory_1998,fischetti_master-equation_1999,knezevic_time-dependent_2013}. These early developments have evolved into a framework that describes reservoirs in terms of broadened modes~\cite{imamoglu_stochastic_1994,garraway_decay_1997,garraway_nonperturbative_1997,zwolak_dynamics_2008}. Within this framework, non--equilibrium Green's functions yield the exact, formal solution for both non--interacting and interacting many--body systems~\cite{gruss_landauers_2016,elenewski_communication_2017,gruss_communication_2017,zwolak_comment_2020,zwolak_analytic_2020}. These results provide a foundation to the overall approach but are limited for \emph{practical} many--body calculations. Consequently, most simulations have been for transport though non--interacting  systems~\cite{sanchez_molecular_2006,subotnik_nonequilibrium_2009,dzhioev_super-fermion_2011,ajisaka_nonequlibrium_2012,ajisaka_nonequilibrium_2013,zelovich_state_2014, zelovich_moleculelead_2015, schwarz_lindblad-driven_2016, zelovich_driven_2016, hod_driven_2016, zelovich_parameter-free_2017, chiang_quantum_2020} (summarized in  Ref.~\cite{elenewski_communication_2017}) or classical thermal energy propagation~\cite{velizhanin_driving_2011, chien_tunable_2013,velizhanin_crossover_2015,chien_thermal_2017, chien_topological_2018}.  However, recent developments in tensor networks deliver a general strategy for many--body calculations within the ERA and related methods.  These methods have enabled large--scale simulations of many--body impurity systems~\cite{wojtowicz_open-system_2020,brenes_tensor-network_2020,lotem_renormalized_2020}, delivered solvers for  dynamical mean--field theory~\cite{arrigoni_nonequilibrium_2013, dorda_auxiliary_2014, dorda_auxiliary_2015, dorda_optimized_2017, chen_Markovian_2019,fugger_nonequilibrium_2020}, and offered new perspectives for open system dynamics~\cite{purkayastha_periodically_2021}.  The theory of extended reservoirs is thus extensive.

\begin{figure}
\includegraphics[width=\columnwidth]{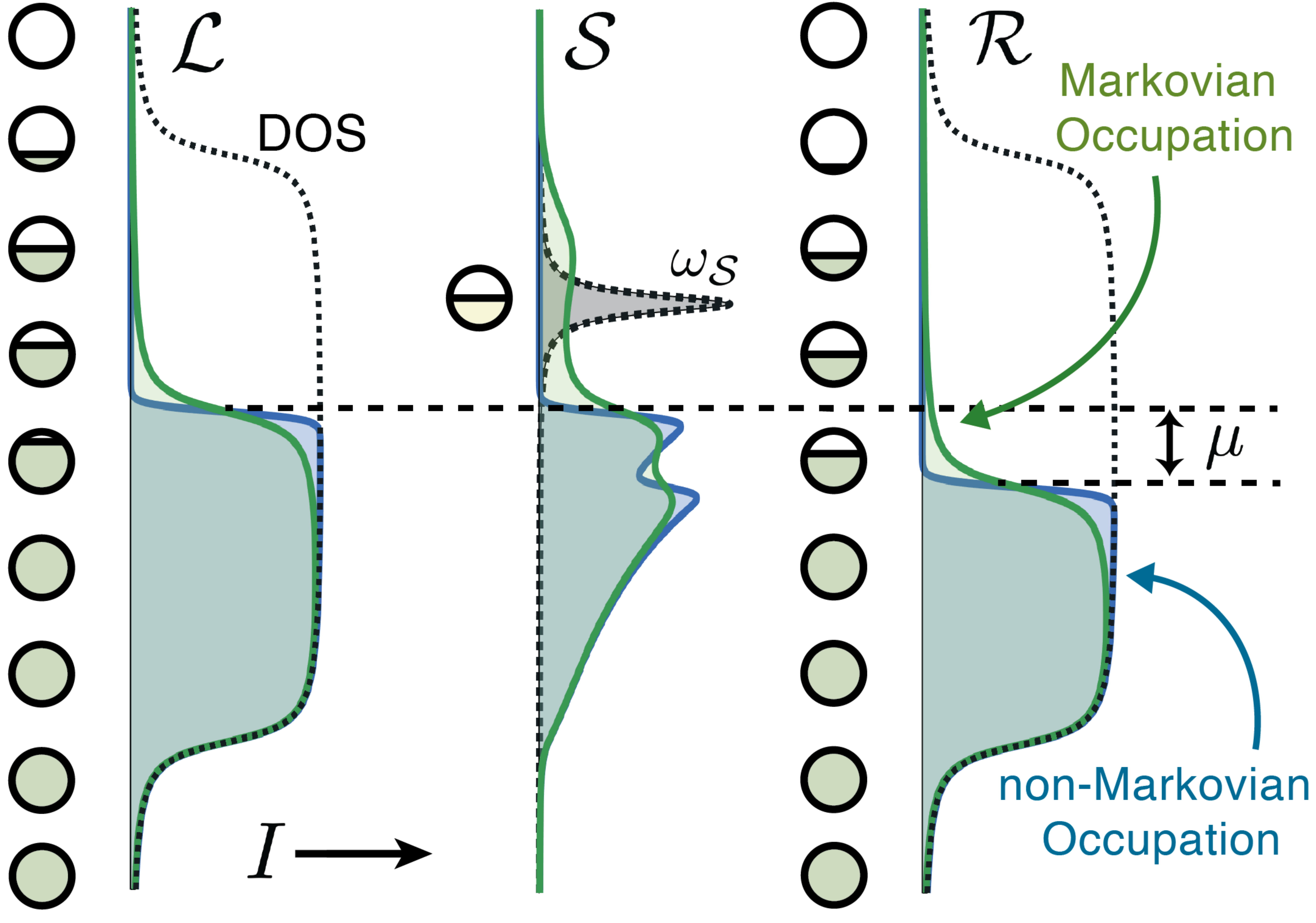}
\caption{{\bf Extended reservoir occupied density of states.} 
The extended reservoir approach represents the left ($\ql$) and right ($\qr$) reservoirs of a device $\qs$ using a finite collection of modes.  An external environment broadens these modes through Markovian or non--Markovian relaxation, which we depict using the occupied density of states for thermodynamic limit (green or blue, respectively).  Transport is then driven by  a chemical potential  difference $\mu$ across $\ql/\qr$, affording a current $I$ through $\qs$. We show a single mode $\omega_\qs$ in isolation (gray, dotted line assuming a small broadening) and depict its occupied DOS when hybridized with Markovian or non--Markovian reservoirs (green or blue, respectively).  Practical calculations generally require Markovian relaxation.  This can anomalously broaden the populated DOS, necessitating judicious parametrization.  Data are from the single-site impurity model in Eq.~\eqref{eq:H1site} with $\omega_\qs = \omega_0$,  system-reservoir hopping $v = \omega_0/8$, $N_\qw = 128$ equally spaced reservoir modes, and $\gamma = \omega_0 / 5$, lying on the low--$\gamma$ side of a Markovian anomaly.
\label{fig:ExtResDOS} }
\end{figure}

The ERA yields a simulation analog of Kramers' turnover~\cite{gruss_landauers_2016,elenewski_communication_2017,gruss_communication_2017,zwolak_analytic_2020,zwolak_comment_2020}, where different transport behaviors are regulated by the strength of relaxation.  This  leads to ``friction--controlled'' currents at very weak (contact--limited) and strong (overdamped) relaxation, with the physically relevant conductance moderating the current at some point between these limits. Here, we go beyond prior developments and demonstrate that additional transport regimes are unveiled for strong impurity scattering. One of these is a \emph{virtual anomaly} associated with tunneling processes and the other a \emph{Markovian anomaly} that emerges from the unphysical Markovian broadening of the occupied density of states (DOS).  These regimes become pronounced at weak system--reservoir coupling, revealing very large currents relative to the continuum limit (i.e., the physically relevant current given by the Landauer or Meir--Wingreen expressions for non--interacting and many--body impurities, respectively). Additional features can emerge due to unrelated processes (e.g., strongly off--resonant tunneling) but the five regimes we discuss appear to be universal, persisting even for weak scattering~\cite{wojtowicz_open-system_2020} and are enhanced when destructive interference is present in the impurity~\cite{chiang_quantum_2020}.

\section{Background}
\label{sec:Back}

We will focus on transport through a central system $\qs$ driven by a chemical potential or temperature drop across non--interacting left ($\ql$) and right ($\qr$) reservoirs, see (Fig.~\ref{fig:ExtResDOS})~\cite{caroli_direct_1971,meir_landauer_1992,jauho_time-dependent_1994}. The full Hamiltonian is 
\[ \label{eq:TotalHam}
H = H_\qs + H_\ql + H_\qr + H_\qi ,
\]
where the system Hamiltonian $H_\qs$ may contain many--body interactions, including electron--electron, electron--photon, and electron--vibration couplings. Furthermore, the reservoir Hamiltonians $H_{\ql (\qr)} = \sum_{k\in \ql (\qr)} \hbar \omega_k \cid{k} \ci{k}$ describe a collection of explicit, non--interacting modes of frequencies $\omega_k$ ($\hbar$ is the reduced Planck's constant). A quadratic Hamiltonian, $H_\qi = \sum_{k\in \ql\qr} \sum_{i\in \qs} \hbar (v_{ki} \cid{k} \ci{i} + v_{ik} \cid{i} \ci{k})$, couples $\qs$ to $\ql\qr$ with strength $v_{ik}=v_{ki}^*$, where $\ci{m}$ ($\cid{m}$) are fermionic annihilation (creation) operators for a level $m \in \ql\qs\qr$. We take the index of sites in $\ql\qs\qr$ to include all relevant labels (state, spin, reservoir or system, etc.)~and sum over all states in a given region.

Finite reservoirs only support a stationary current when external reservoirs are present. Thus, we will solve the (Markovian) Lindblad master equation 
\begin{align} \label{eq:fullMaster}
\dot{\rho} = - \frac{\im}{\hbar} [H, \rho]
    &+ \sum_{k\in\ql\qr} \gamma_{k+} \left( \cid{k} \rho \ci{k}
        - \frac{1}{2} \left \{ \ci{k} \cid{k}, \rho\right \}\right) \notag \\
    &+ \sum_{k\in\ql\qr} \gamma_{k-} \left( \ci{k} \rho \cid{k}
        - \frac{1}{2} \left \{ \cid{k} \ci{k}, \rho \right \} \right)
\end{align}
for the $\ql \qs \qr$ system with Markovian relaxation in $\ql\qr$ ($\{\cdot,\cdot\}$ is the anticommutator) and a corresponding scenario for non--Markovian relaxation. Throughout this work, we use the term Markovian relaxation when it follows from a time--local Lindblad master equation as in Eq.~\eqref{eq:fullMaster}. Non--Markovian relaxation corresponds to evolution with similar retarded and advanced Green's functions, but instead with proper occupations as described in Sec.~\ref{sec:Current} --- that would, however, follow from an inherently non--time--local master equation. Ref.~\cite{gruss_landauers_2016} shows the Hamiltonian for the $\ql\qs\qr$ system and environment that gives rise to the non--Markovian equation of motion when the environment is integrated out.

The first term in Eq.~\eqref{eq:fullMaster} gives the evolution of the full, many--body density matrix $\rho$ according to the Hamiltonian $H$ in Eq.~\eqref{eq:TotalHam}.  Open dynamics arise from the Lindbladian terms, which inject or deplete particles to or from the modes $k$ at rates $\gamma_{k+}$ and $\gamma_{k-}$, respectively. If we adopt a convention where these rates are  $\gamma_{k+} \equiv \gamma_k f^\alpha (\omega_k)$ and $\gamma_{k-} \equiv \gamma_k [1 - f^\alpha (\omega_k)]$, the $\ql\qr$ reservoir modes will relax to an equilibrium Fermi--Dirac distribution  $f^\alpha (\omega_k)$ (with $\alpha \in \{\ql,\qr\}$) when decoupled from $\qs$. The chemical potential  $\mu_\alpha$ of each reservoir is included in $f^\alpha (\omega_k)$, where we take $\mu_\ql = -\mu_\qr = \mu/2$ for the potential bias between reservoirs $\mu$. The result is a pseudo--equilibrium state,  as it does not incorporate relaxation--induced broadening of the extended reservoirs' modes.  We will show how this leads to a transport anomaly at moderate to strong $\gamma$.

Reference~\cite{gruss_landauers_2016} provides the formal solution for the  steady--state current corresponding to Eq.~\eqref{eq:fullMaster}, which is valid both for interacting and non--interacting many--body systems $\qs$ (see Refs.~\cite{zwolak_analytic_2020,zwolak_comment_2020} for a unified derivation and fully analytic solutions for proportional coupling). It was also proven that the formal solution limits to either the Meir--Wingreen expression or the Landauer formula. These expressions require us to find the single--particle Green's function in a system with many--body interactions. This can be complicated and often requires many approximations.  Alternatively, one can solve Eq.~\eqref{eq:fullMaster} numerically using established techniques such as tensor networks (as recently applied to transport in Refs.~\cite{wojtowicz_open-system_2020,brenes_tensor-network_2020,lotem_renormalized_2020}). 

We begin by addressing systems that are proportionally coupled to the electrodes. This requires identical distributions of mode frequencies and relaxation strengths in each finite, extended reservoir. The distribution of the system--reservoir coupling constants must also be the same up to some overall proportionality constant.  We will show that if we apply a small level shift between reservoirs, we can effectively ``turn off'' the virtual anomaly. This general observation will carry over to non--proportionally coupled and interacting setups. Even if the continuum limit has such a symmetry, there can be finite representations that limit to proportional coupling yet have it formally broken. In fact, breaking proportional coupling and using non--Markovian relaxation results in a three--regime Kramer turnover, which is more well behaved, with both anomalies removed.~\footnote{Breaking proportional coupling is much more reflective of physical reality, where defects and other inhomogeneities in structure will break such a strong symmetry.}

\subsection{Steady--state current}
\label{sec:Current}

The steady--state current associated with Eq.~\eqref{eq:fullMaster} has an exact solution for arbitrary non--interacting systems and reservoirs~\cite{gruss_landauers_2016,zwolak_analytic_2020,zwolak_comment_2020} 
\[ \label{eq:nonintCurr}
I = e \int\frac{d\omega}{2\pi} \tr \left[ \tilde{\bGa}^\ql \bG^a \bGa^\qr \bG^r - \bGa^\ql \bG^r \tilde{\bGa}^\qr \bG^a \right] ,
\]
encompassing both Markovian and non--Markovian relaxation. Here, $e$ is the electron charge, $\bG^{r(a)}$ is the retarded (advanced) Green's function of the full system, and ${\bGa}^{\ql(\qr)}\, (\tilde{\bGa}^{\ql(\qr)}$) are the unweighted (weighted) spectral functions (we note to the reader that in some contexts the spectral function is also known as a ``hybridization function''), which all depend on the frequency $\omega$. We employ notation where bold symbols indicate matrices with standard matrix multiplication assumed. For non--Markovian (nM) relaxation, this formula reduces to 
\[ \label{eq:nonintCurrStandard}
I \overset{\nM}{=} e \int\frac{d\omega}{2\pi} \left( f^\ql (\omega) - f^\qr (\omega) \right) \tr \left[ \bGa^\ql \bG^r \bGa^\qr \bG^a \right] ,
\]
where we indicate specific conditions for the equation by labeling the equality (i.e., here ``nM'' indicates that this expression is for non--Markovian relaxation). The explicit forms for the underlying Green's functions are $\bG^{r (a)} = 1/(\omega - \bHS - \bS^{r (a)})$ with self--energies $\bS^{r (a)}=\sum_{k\in\ql\qr} g^{r (a)}_k \ok$. These expressions are identical for Markovian and non--Markovian relaxation. We use $\kk$ to denote the coupling vector between mode $k\in\ql\qr$ and all sites $i\in\qs$, i.e., $\inner{i}{v_k}=v_{ik}$, and write $g_{k}^{r(a)}=1/(\omega-\omega_k \pm \im \gamma_k/2)$ for the retarded (advanced) Green's function with $k\in\ql\qr$. These latter quantities have $\gamma_k>0$ but are isolated from the system. The single--particle Hamiltonian describing $\qs$ is $\bHS$, where $H_\qs = \sum_{i,j\in\qs} [\bHS]_{i,j} \cid{i} \ci{j}$. General and exact results for the steady--state current with many--body impurities, in the presence of Markovian and non--Markovian relaxation, can be found in Refs.~\cite{gruss_landauers_2016,zwolak_analytic_2020}. We study many--body systems numerically, and provide analytic results for non--interacting systems.

As a final component, one also needs the spectral density  $\bGa^{\ql (\qr)} = \im ( \bS^r_{\ql (\qr)} - \bS^a_{\ql (\qr)}) = -2 \Im \bS^r_{\ql (\qr)}$, 
\[  \label{eq:unweightedsigma_gen}
\bGa^{\ql(\qr)}(\omega) = \im\sum_{k\in \ql(\qr)} 
                       \left[g_{k}^{r}(\omega) - g_{k}^{a}(\omega) \right] \ok.
\]
The population--weighted counterpart is
 \[  \label{eq:weightedsigma_gen}
\tilde{\bGa}^{\ql(\qr)}(\omega) = \im\sum_{k\in \ql(\qr)} \tilde{f}_k
                       \left[g_{k}^{r}(\omega) - g_{k}^{a}(\omega) \right] \ok.
\]
Based on this, Markovian and non--Markovian relaxation only differ in how we evaluate the Fermi--Dirac occupations, $\tilde{f}_k$,
\[
\tilde{f}_k=\begin{cases}
f^{\alpha} \left(\omega\right) & \textbf{non--Markovian}\\
f^{\alpha} \left(\omega_{k}\right) & \textbf{Markovian} 
\end{cases} ,
\]
for reservoir $\alpha \in \{\ql, \qr\}$ and $f^{\alpha} (\omega)$ having bias $\mu_{\alpha}$. These distributions set mode occupations to an inherently unphysical Markovian equilibrium or to a physical non--Markovian equilibrium. The latter occupies modes to give an appropriate broadening and thus gives a proper Fermi level as shown in Fig.~\ref{fig:ExtResDOS}.

For systems that are proportionally coupled yet otherwise arbitrary (e.g., in structure, with or without many--body interactions, and for Markovian or non--Markovian relaxation), the steady--state current is~\cite{zwolak_comment_2020,zwolak_analytic_2020}
\[ \label{eq:CurrPC}
I \overset{\PC}{=} \im e \frac{\lambda}{1+\lambda} \int\frac{d\omega}{2\pi}\, \tr \left[ \Delta \tilde{\bGa} \left\{ \bG^{r} - \bG^{a} \right\} \right],
\]
where we use label ``$\PC$" to indicate proportional coupling.
Stated formally, this means that~\cite{meir_landauer_1992,jauho_time-dependent_1994}
\[
\bGa^\qr = \lambda \bGa^\ql \equiv \lambda \bGa,
\] 
for some positive constant $\lambda$.  The current $I = \frac{\lambda}{1+\lambda} I_\ql + \frac{1}{1+\lambda} I_\qr$ is then an average over the left, $I_\ql$, and right, $I_\qr$, currents to/from $\qs$. The current in Eq.~\eqref{eq:CurrPC} also contains a difference in weighted spectral densities
\[ \label{eq:DiffSpectral}
\Delta \tilde{\bGa} \overset{\PC}{=} \im \sum_{k\in \ql} \left(\tilde{f}_k^\ql - \tilde{f}_k^\qr \right) \left[g_{k}^{r}(\omega) - g_{k}^{a}(\omega) \right] \ok ,
\]
where the sum only runs over states $k$ in the left reservoir.  As per the proportional coupling requirements, the left and right reservoirs are equivalent in their mode placement and relaxation, with couplings related by $v_{k^\p i}= v_{ki} \sqrt{\lambda} \delta_{kk^\p}$ with $k^\p \in \qr$, $k \in \ql$. The factor of $\lambda$ does not appear in Eq.~\eqref{eq:DiffSpectral}, as averaging brings it out front in Eq.~\eqref{eq:CurrPC}.

We may simplify this further for Markovian (M) relaxation. The occupation factors in Eq.~\eqref{eq:CurrPC} are then evaluated at $\omega_k$, removing them from the $\omega$ integration and giving an analytic expression \cite{zwolak_analytic_2020,zwolak_comment_2020}
\[ \label{eq:GenCurr_PC}
I \overset{\PC, \M}{=} \frac{-2 e \lambda}{1+\lambda} \sum_{k\in \ql} \left(\tilde{f}_k^\ql - \tilde{f}_k^\qr \right) \bk \Im \left[{\bG^r (\omega_k + \im \gamma_k/2 )}\right] \kk .
\]
The integral in Eq.~\eqref{eq:CurrPC} cannot be evaluated for the non--Markovian case due to the appearance of $f^\alpha(\omega)$.

\subsection{Kramers' turnover}
\label{sec:Kramers}

In the presence of relaxation, particle and thermal transport yield behavior analogous to Kramers' turnover for condensed--phase chemical reaction rates~\cite{gruss_landauers_2016, velizhanin_crossover_2015}. Stated succinctly, Kramers' problem describes reactants that must overcome a free--energy barrier to become products, while also being subject to friction and noise due to the encapsulating solvent~\cite{kramers_brownian_1940,hanggi_reaction-rate_1990}.  When friction is weak, the reaction rate will be linearly proportional to the strength of this friction. This quantity defines the rate at which equilibrium is reestablished, restoring the proportion of reactants that possess sufficient free energy to overcome the barrier.  For large friction, this restoration process is rapid.  However,  strong friction can localize reactants in the initial state, decreasing the reaction rate in inverse proportion to the frictional strength. Between these limits there is a region where the intrinsic, transition state rate is dominant.

A similar phenomenon occurs for transport simulations.  At weak relaxation (friction, noise), the current is limited by coupling to the external (implicit) environments, which control the rate of particle injection and depletion in the reduced $\ql \qs \qr$ system. This regime is characterized by a  current~\cite{gruss_landauers_2016,zwolak_analytic_2020,zwolak_comment_2020}
\[ \label{eq:Curr_WG_PC}
I \overset{\PC, \M}{\approx} \frac{2e \lambda}{(1+\lambda)^2} \sum_{k \in \ql} \gamma_k \left(\tilde{f}_k^\ql - \tilde{f}_k^\qr \right),
\] where factors $\gamma_k$ contribute proportionally.

Conversely, at strong relaxation, coherence with the central region is destroyed, localizing particles in the extended reservoir modes. This results in the current~\cite{gruss_landauers_2016,zwolak_analytic_2020,zwolak_comment_2020} 
\[ \label{eq:Curr_SG_PC}
I \overset{\PC, \M}{\approx} \frac{4 e \lambda}{1+\lambda} \sum_{k\in \ql} \frac{\ik}{\gamma_k} \left(\tilde{f}_k^\ql - \tilde{f}_k^\qr \right) ,
\]
with $\ik=\inner{v_k}{v_k}$, and the factors $\gamma_k$ giving inversely proportional contributions.

Equations~\eqref{eq:Curr_WG_PC} and~\eqref{eq:Curr_SG_PC} give the asymptotic limits of Eq.~\eqref{eq:GenCurr_PC} for Markovian relaxation and proportional coupling, with a linear regime for weak relaxation and an inverse regime at strong. 
The primary objective of this paper is to extend prior developments, Ref.~\cite{gruss_landauers_2016}, by accounting for additional regimes that occur for for weak coupling and  non-resonant conditions. We will show that these linear and inverse regimes flank a pair of anomalous transport regimes. These anomalous regimes, in turn, flank the physical regime corresponding to continuum reservoirs (where relaxation is not present).

\subsection{Example models}
\label{sec:Model}

For demonstrative purposes, we will focus on two central systems $\qs$: a single-site (non--interacting) impurity and a two--site (interacting) impurity. For the single-site case,
\begin{equation}
\label{eq:H1site}
H_\qs = \hbar \, \omega_\qs \cid{1} \ci{1},
\end{equation}with on--site mode frequency $\omega_\qs$. As a representative many--body model, we take
\begin{equation}
    \label{eq:H2sites}
    H_\qs = \hbar \, v_\qs (\cid{1} \ci{2} + \cid{2} \ci{1} ) + \hbar \, U n_{1} n_{2},
\end{equation}where the hopping frequency between sites is $v_\qs$, the number operator is $n_j=\cid{j}\ci{j}$,  the density--density interaction strength is $U$, and on-site mode frequencies are fixed at zero. 

We consider a setup, where the central region $\qs$ is situated between spatially one--dimensional (1D) reservoirs i.e., one--dimensional chains with hopping of frequency $\omega_0$ between nearest--neighbor lattice sites.  The last site of each reservoir is attached to a single site of $\qs$ via hopping frequency $v$. This makes for a unique arrangement in the single--site model. 

For the two-site model, we address two variants of  attachment. First, we consider the destructive interference model of Ref.~\cite{chiang_quantum_2020} (but with a many--body interaction $U$), where the first site is connected to both reservoirs and the second to neither (i.e., it is proportionally coupled). In the second variant, the sites form a serial junction with the first site connected only to $\ql$ and the second only to $\qr$, as in Ref.~\cite{wojtowicz_open-system_2020}. There is no proportional coupling in this case since each site is coupled to a single reservoir.

In the continuum limit, when each reservoir chain extends to infinity, each of our reservoirs contributes to retarded self--energy of the system site it is attached to so that (see, e.g., Ref.~\cite{gruss_landauers_2016})
\[ \label{eq:SECont}
\bS^r = \frac{8 v^2}{W^2} \left( \omega + \im \frac{\gamma}{2} - \im \sqrt{\frac{W^2}{4} - \left( \omega + \im \frac{\gamma}{2} \right)^2} \right),
\]
where $W=4\omega_0$ is the reservoir bandwidth. Equation~\eqref{eq:SECont} is sufficient to derive the Green's functions and spectral densities. This choice will minimally impact most of the results. This also holds for the performance of numerical simulations, where we employ the reservoirs' energy bases instead of their chain representations.  Nonetheless, our numerical results will be specific to this setup and some equations will be specific to the 1D DOS. 

The reservoirs can be discretized in a number of ways, provided that they reproduce the continuum in the asymptotic limit.  We will present equations for ``evenly spaced'' (ES) reservoir modes as well as ``transformed spacing'' ($\OD$). 
$\OD$ is defined by the eigenfrequencies of a finite, spatially one--dimensional 
chain of~$N_\qw$~sites. These can be found using the canonical transformation $c_k=\sum_{j\in\alpha}\qu^\dg_{kj} c_j$ with $\qu^\dg_{kj}=\sqrt{2/\left(N_\qw+1\right)} \sin\left(jk\pi/\left(N_\qw+1\right)\right)$ and $k=1,\ldots,N_\qw$. This yields frequencies $\omega_{k}= 2\omega_0\cos\left(k\pi/\left(N_\qw+1\right)\right)$ and couplings $v_{ki}=v_{k}= v\sqrt{2/(N_\qw+1)}\sin\left(k\pi/\left(N_\qw+1\right)\right)$. We note that $k$ carries both a numerical index and a reservoir index $\alpha$. This discretization gives the system--reservoir coupling at the Fermi level ($\hbar\omega_F = 0$)
\[
\vFs \overset{\OD}{=} \vs \Delta_F / (\pi \omega_0),
\]
as well as the level spacing
\[
\Delta_F = 2 \omega_0 \pi / (N_\qw+1).
\]
These quantities will be used below. Similar expressions hold for the evenly spaced discretization, where we have 
\[
\vFs \overset{\ES}{=} \vs \Delta_0 /(\pi \omega_0),
\]
and  
\[
\Delta_0 = W/N_\qw = 4\omega_0/N_\qw ,
\]
where reservoir modes $\omega_k$ are placed in midpoints of each of the $N_\qw$ bins of size $\Delta_0$ (thus filling in the bandwidth). We employ the midpoint approximation to set coupling constants to the reservoirs in our MPS calculations, and ensure that they properly limit to the continuum, i.e., $v_{kj} = 2 v [ \Delta_k/ (\pi W) \sqrt{1- (2\omega_k / W)^2} ]^{1/2}$ for the interval $\omega_k \pm \Delta_0/2$. This coupling method performs comparably to the integrated coupling for common reservoir discretization methods~\cite{elenewski_performance_2021}, where one integrates the spectral density over a symmetric frequency range $\omega_k \pm \Delta_0/2$ about the finite reservoir mode $\omega_k$.  This integrated quantity allows one to define an equivalent coupling for $\omega_k$ that maintains the total weight from the continuum reservoir in this spectral region,  given by $v_{kj} = v \pi^{1/2} [K(\omega_k + \Delta_k/2) - K(\omega_k - \Delta_k/2)]^{1/2}$  where $K(\omega) = 2 \omega/W (1 - 4\omega^2 / W^2)^{1/2} + \csc^{-1} (W / (2\omega))$.  We use these integrated couplings for our exact, non--interacting calculations.

\section{Results}
\label{sec:Results}
Figure~\ref{fig:ExtResDOS} depicts a typical extended reservoir and its weighted spectral functions (i.e., the occupied density of states in the reservoirs). The DOS and occupied DOS are taken from continuum limit expressions for a single site coupled to a pair of one--dimensional reservoirs (transformed to the energy basis).  It is clear that non--Markovian relaxation results in an actual Fermi level, while Markovian relaxation gives unphysical broadening \footnote{We note that this unphysical behavior specifically refers to applicability when describing fermions, as Lindblad master equations always give a mathematically proper quantum evolution.}. This broadening is responsible for the anomaly observed at  moderate--to--strong relaxation strength, as well as for the zero--bias currents associated with asymmetric (non--proportionally coupled) reservoirs~\cite{gruss_landauers_2016}.

\begin{figure}[t]
\includegraphics[width=\columnwidth]{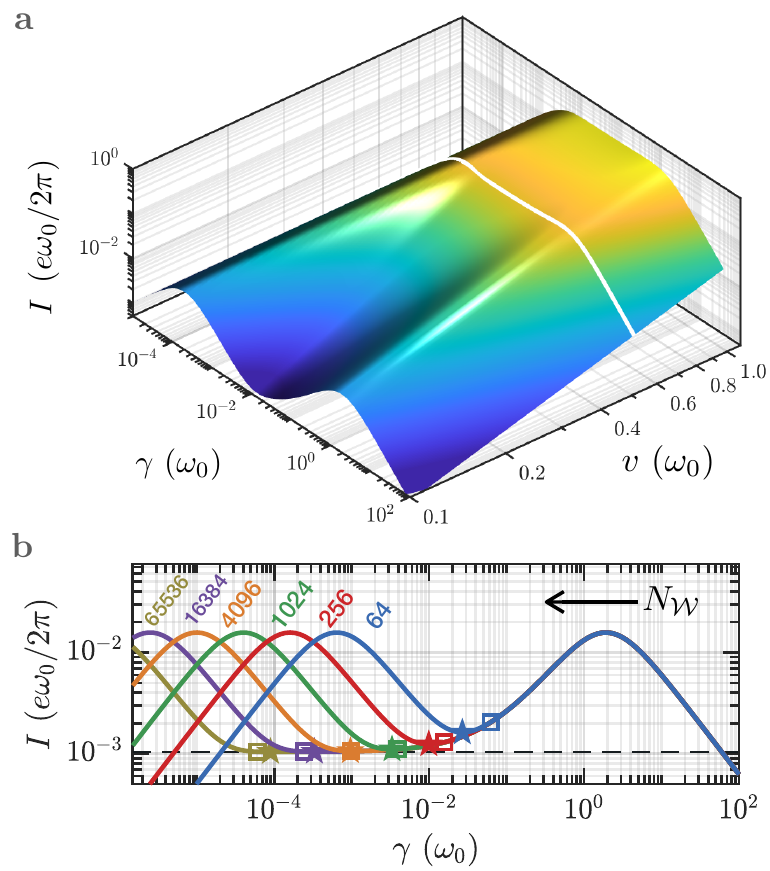}
\caption{{\bf Anomalous currents and Kramers' turnover for a single--site impurity.} (a) Kramers' turnover versus relaxation $\gamma$ and  coupling $v$. Strong scattering (weak coupling) reveals two regimes: a ``virtual anomaly'' for weak--to--moderate relaxation and a ``Markovian anomaly'' for moderate--to--strong relaxation. The former is due to resonant $\ql,\qr$ modes which artificially increase particle currents. The respective virtual transitions through $\qs$ are eventually overdamped and turn over. The Markovian anomaly is due to unphysical smearing of the populated DOS by the relaxation, again increasing the current. This too is ultimately overdamped at stronger relaxation. These regimes flank the intermediate $\gamma$ regime where the continuum limit is best recovered. Results correspond to a single-site impurity in Eq.~\eqref{eq:H1site} with $\omega_S=\omega_0$, and $N_\qw=128$ equally--spaced reservoir modes. Couplings are given by the (integrated) strength within each mode bin and the bias is $\mu=\omega_0/2=2\mu_\ql=-2\mu_\qr$.  The white line designates the onset of the anomaly parameter space, according to Eq.~\eqref{eq:Ratio}. (b) Kramers' turnover at weak coupling versus $\gamma$ for several $N_\qw$. On the logarithm scale, the size of the continuum regime is proportional to $\log N_\qw$, while anomalies slow its convergence to the Landauer limit (dashed black line). Parameters are the same as in (a) but at   $v=\omega_0/8$ and for various $N_\qw$.  Stars denote the optimal relaxation $\gamma^\star$ while squares correspond to $\gamma^\Hsquare$ (a popular choice; proportional to the level spacing in the reservoirs), demonstrating a shift of $\gamma^\Hsquare$ from the optimum. }\label{fig:AnomalousCurrents} 
\end{figure}

Figure~\ref{fig:AnomalousCurrents} shows the steady--state current for Markovian relaxation, Eq.~\eqref{eq:GenCurr_PC}, for the single--site model in Eq.~\eqref{eq:H1site}. When the coupling is strong ($v \approx \omega_0$), the current assumes a well--defined plateau as a function of the relaxation rate $\gamma$. We take $\gamma$ to be the same for all reservoir modes.  It is here that the steady--state current generically reproduces the continuum reservoir limit.  This plateau is markedly different at weak coupling ($v \ll \omega_0$), where large features exist at moderately small and moderately large $\gamma$ (the peaks of which mark the plateau edges at strong coupling).  Situated between these features is a regime that corresponds to the continuum limit (Landauer's formula for non--interacting cases), a region that increases in size for larger reservoirs.

We will address the origin of these anomalies in linear response, as the resulting expressions are easily interpreted (the same underlying phenomena occur out of linear response). Notably, the non--Markovian and Markovian relaxation have different forms, yielding distinct implications. Starting from Eq.~\eqref{eq:CurrPC}, the current for non--Markovian relaxation is 
\[ \label{eq:CurrNMLR}
I \LR{\overset{\PC, \nM}{\approx}} -\frac{e \mu}{\pi}  \frac{\lambda}{1+\lambda} \left. \tr \left[ \bGa \Im \bG^{r} \right] \right|_{\omega=0},
\]
where $\mu = \mu_\ql - \mu_\qr$ is a symmetrically applied bias, resulting in evaluation at $\omega=0$ (although it can be evaluated anywhere in the suitably small bias window).  Here, we  have also replaced the difference in the weighted spectral density with  the spectral density {of $\ql$}. In contrast, Markovian relaxation gives
\[ \label{eq:CurrMLR}
I  \LR{\overset{\PC, \M}{\approx}} -\frac{e N_\qb}{\pi} \frac{\lambda}{1+\lambda} \int d\omega \, \tr \left[ \bGa_F  \Im \bG^{r} \right] .
\]
The integral is the key difference: this gives an overlap of the system DOS at all frequencies with the spectral density, while  taking only contributions from the Fermi level modes (taken at $\omega_k = 0$ and designated $F$),
\[
\bGa_F  = \im \left[ g_{F}^{r}(\omega) - g_{F}^{a}(\omega)  \right] \oF .
\]
Alternatively, we can start with the analytic solution~\eqref{eq:GenCurr_PC} for Markovian relaxation, yielding
\[ \label{eq:CurrMLRa}
I \LR{\overset{\PC, \M}{\approx}} - 2 e N_\qb \frac{\lambda}{1+\lambda} \bF \Im \bG^r_F \kF ,
\]
which is equivalent to Eq.~\eqref{eq:CurrMLR}. The Green's function,
\[ \label{eq:GrF}
\bG^r_F = \bG^r(0 + \im \gamma_F/2 ) ,
\]
is at the Fermi level (assuming a smooth $\gamma_k$ there) and
\[ \label{eq:Nmu}
N_\qb = \sum_{k\in\ql} \tilde{f}_k^\ql - \tilde{f}_k^\qr 
\approx
\begin{cases}
\begin{array}{c}
2 N_\qw \frac{\mu}{\pi W} \\
N_\qw\frac{\mu}{W}
\end{array} & \begin{array}{c}
\text{for} \,\, \OD,\\
\text{for} \,\, \ES
\end{array}\end{cases}
\]
is the effective number of modes in the bias window, which is typically not an integer (though it will be  at zero temperature and for particular arrangements of modes).  Linear response requires that $\tilde{f}_k^\ql - \tilde{f}_k^\qr$ is negligible except around the Fermi level or, more broadly, non--negligible only in the region where there is little variation for the other factors present in Eq.~\eqref{eq:CurrMLRa}. The approximate equality in Eq.~\eqref{eq:Nmu} reflects that the level spacing varies in the bias window and that modes can cover more than just the bias window (e.g., $\mu/\Delta_F$ is not always an integer).~\footnote{For closed systems, one will get systematic errors when the mode ``bins'' do not exactly fill the bias window~\cite{zwolak_communication_2018}.}

The non--Markovian and Markovian expressions~\eqref{eq:CurrNMLR} and~\eqref{eq:CurrMLR}, converge when relaxation is weak. The difference between these is nonetheless pronounced at strong relaxation, where non--Markovian effects give an overlap between the spectral density and system DOS around the Fermi level.  Conversely, Markovian relaxation gives an overlap over all frequencies. This issue originates in the violation of the fluctuation--dissipation theorem and has important consequences.

We will examine this behavior in detail for a single--site impurity. Its Green's function is 
\[ \label{eq:Gr1S}
\bG^r \overset{\PC}{=} \frac{1}{\omega-\omega_\qs - (1+\lambda)\sum_{k\in\ql} \frac{ \left| v_k \right|^2 }{\omega-\omega_k+\im \gamma_k/2}} ,
\]
where $\omega_\qs$ is {on-site frequency for the impurity} and $v_k$ is a scalar since there is {only} one site {in $\qs$}. In the following, we will take equally spaced reservoir modes that are finite realizations of a one--dimensional lattice with real--space hopping frequency $\omega_0$.  Nonetheless, all of our analytic results will apply to arbitrary, proportionally--coupled reservoirs (see Sec.~\ref{sec:Model}.
For simplicity we assume a homogeneous relaxation  rate, i.e., $\gamma_k =\gamma$.

Employing Eq.~\eqref{eq:Gr1S} in Eq.~\eqref{eq:CurrMLRa} allows us to interpolate between all regimes for weak system--reservoir coupling ($\WC$). At very small $\gamma$, the contribution from the self--energy becomes dominant and $\bG^r$ scales proportional to  $\gamma$, yielding $I \propto \gamma$. Conversely, for very large $\gamma$, the self--energy contribution is small. However, since Eq.~\eqref{eq:CurrMLRa} is evaluated at $\omega_k + \imath \gamma/2$, the dominant factor in $\bG^r$ is inversely proportional to  relaxation rate leading to $I\propto 1/\gamma$. Approximating the self--energy with the dominant contribution at small $\gamma$, the current is 
\[ \label{eq:CurrApp}
I \LR{\overset{\PC, \M, \WC}{\approx}} \frac{2 e \lambda N_\qb \vFs}{1+\lambda} \frac{\gamma/2 + (1+\lambda) \vFs/\gamma}{\omega_\qs^2 + \left(\gamma/2 + (1+\lambda) \vFs/\gamma \right)^2 } ,
\]
where the reservoir size is embedded in both $N_\qb$ [see Eq.~\eqref{eq:Nmu}] and $\vFs \propto v^2/N_\qw$. This approximate interpolation breaks down at large system--reservoir coupling but becomes exact as the coupling goes to zero for off--resonant tunneling (e.g., $\omega_\qs$ outside of the bias window). It nonetheless captures the physics of different turnover regimes, as well as their quantitative behavior when coupling is weak. For finite reservoirs, this expression converges to the exact {\em finite reservoir} result, not the Landauer formula, as $\vs/\omega_\qs$ drops below the finite level spacing.  Here, $v$ is the total, real--space coupling and not $v_k$ which already depends on $1/\sqrt{N_\qw}$. Due to this, the order of limits is important, as we will discuss later.

As a point of reference, we introduce $I^\circ$ to denote the continuum (Landauer) result, 
\[ \label{eq:CurrLand}
I^\circ \LR{\overset{\PC}{\approx}} \frac{e}{2\pi} \frac{4 \lambda \left| v \right|^4}{(1+\lambda)^2 \left| v \right|^4 +\omega_0^2 \omega_\qs^2} \mu .
\]
This expression is given in linear response. Nonetheless, the correspondence with the extended reservoir approach also becomes exact for non--linear response and arbitrary $\qs$ as the reservoirs approach the continuum and relaxation limits to zero~\cite{gruss_landauers_2016,zwolak_analytic_2020,zwolak_comment_2020}.

\subsection{Duality between current anomalies}
\label{sec:Duality}

We can isolate specific aspects of turnover by rewriting Eq.~\eqref{eq:CurrApp} in a Lorentzian form for the weak--to--moderate and moderate--to--strong anomalies,
\[ \label{eq:Duality}
I \LR{\overset{\PC, \M, \WC}{\approx}}\frac{2 e \lambda N_\qb \vFs}{1+\lambda} \frac{\tilde{\gamma}}{\omega_\qs^2 + \tilde{\gamma}^2}.
\]
This expression permits an easy separation of relaxation regimes,
\[
\begin{array}{c}
\mathbf{small}\,\gamma\\
\tilde{\gamma} \Rightarrow (1+\lambda)\vFs/\gamma
\end{array} \boldsymbol{\Leftrightarrow} \quad\quad\begin{array}{c}
\mathbf{large}\,\gamma\\
\tilde{\gamma} \Rightarrow \gamma/2
\end{array} \quad ,
\]
which are distinguished by the parameter $\tilde \gamma$~\footnote{While we expect some other  models will show such a strong duality, it will not be general. However, we expect the qualitative discussion to apply beyond the one--site impurity and that some of the quantitative expressions for weak coupling will hold}. We immediately find some useful results. First and foremost, the anomalous current maximum, $I^\top$, is the same for both anomalies, {\em independent of $N_\qw$}, and is given by 
\[ \label{eq:AnomalousCurr}
I^\top \approx \frac{e \lambda N_\qb \vFs}{1+\lambda} \frac{1}{\omega_\qs} \overset{\ES}{\approx} \frac{e \lambda \vs \mu}{(1+\lambda) \pi\omega_0 \omega_\qs}.
\]
The second expression assumes evenly spaced modes (that represent a one--dimensional spectral function). In linear response, this implies that the current anomaly will surpass the Landauer value by a factor of
\[
\frac{I^\top}{I^\circ} \overset{\ES}{=} \frac{\omega_0 \omega_\qs}{2(1+\lambda)\vs}, \label{eq:Ratio}
\]
where the Landauer current $I^\circ$ is in Eq.~\eqref{eq:CurrLand}.  This relation diverges as the coupling decreases, implying that the anomalies progressively dwarf the Landauer current. If we take $I^\top = I^\circ$ at weak coupling, Eq.~\eqref{eq:Ratio} yields the parameter values where the anomalous regime is revealed (see the white turnover curve in Fig.~\ref{fig:AnomalousCurrents}a). 

At weak coupling, the current~\eqref{eq:CurrApp} will not develop a well--defined plateau, but will instead give a distinct minimum between anomalies.  Due to the nature of this arrangement, this minimum will occur at the  {\em geometric} midpoint between the anomalies. This will not hold for strong coupling, as the minimum will be distorted by non--perturbative effects. Note that additional features may appear in this region, such as an additional nonlinear dip for strongly off--resonant tunneling. 

\begin{figure}[t!]
\includegraphics[scale=1.0]{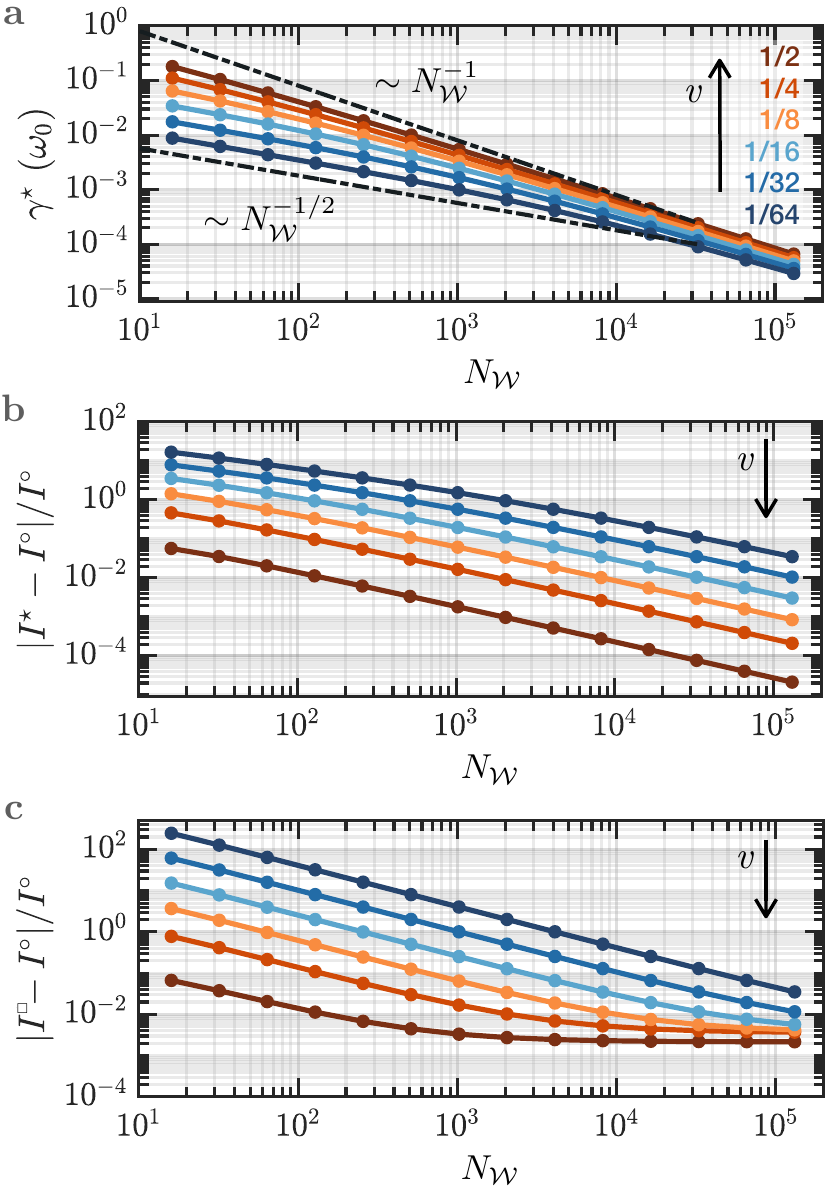}
\caption{{\bf Optimal relaxation and accurate simulation.} Extended reservoir simulations typically employ a relaxation $\gamma$ proportional to the level spacing ($\gamma \propto 1/N_\qw$). However, the dual anomalies, present for Markovian reservoirs and weak coupling, indicate that the optimal estimator should be proportional to the inverse square root of modes ($\gamma \propto 1/\sqrt{N_\qw}$). (a) The actual optimal estimator $\gamma^\star$ (found numerically by comparing turnover to the continuum limit $I^\circ$) transitions from scaling as $1/\sqrt{N_\qw}$ to $ 1/N_\qw$ as the coupling approaches unity.  Deviations are seen for large $N_\qw$ at weak coupling, as explained in the text.  Best fits to $A/N_\qw^p$ scaling give parameters $(A,p) = (1.72,-0.81)$, $(0.87,-0.74)$, $(0.41,-0.69)$, $(0.18,-0.59)$,  $(0.08,-0.55)$, and $(0.04,-0.52)$ for $v$ from large to small (couplings $v$ are labeled in units of $\omega_0$ and increase between curves as denoted by the arrows).  The standard deviation for all fitting parameters is $\pm 0.01$.  (b) Relative error, $\vert I^\star - I^\circ \vert / I^\circ$, versus $N_\qw$ for the current $I^\star=I (\gamma^\star)$ for the $\gamma^\star$ from (a)  compared to the continuum limit value.  (c) Relative error, $\vert I^\Hsquare - I^\circ \vert / I^\circ$, for the estimate $I^\Hsquare = I (\gamma^\Hsquare)$, where the relaxation $\gamma^\Hsquare = W / N_\qw$ is equal to the level spacing. This current estimate misses the plateau by a fixed amount as $N_\qw$ increases. In other words, the estimate rides the edge of the virtual anomaly as $N_\qw \to \infty$ and $\gamma \to 0$ (taken together) and {\em will not converge to the continuum limit}. Model parameters are the same as Fig. \ref{fig:AnomalousCurrents}(b), except where specified.}\label{fig:AccSim}
\end{figure}

The duality, Eq.~\eqref{eq:Duality}, also indicates that the maximum of both anomalies will be found at $\tilde{\gamma} = \omega_\qs$, as defined by the two particular $\tilde{\gamma}$.~\footnote{One can also take the point where the two Lorentzian turnovers intersect, which gives the same value.} The {\em optimal} relaxation $\gamma^\star$ for estimating the current is at their geometric mean
\[ \label{eq:OptGam}
\gamma^\star \overset{\WC}{\approx} \vFv \sqrt{2(1+\lambda)} \overset{\ES}{=}
2\vv \sqrt{\frac{2 (1+\lambda)}{\pi N_\qw}}
\]
in the weak--coupling limit. By this, we mean that it gives the best estimate for the Landauer limit while excluding the ``accidental'' crossings within the $I \propto \gamma$ and the $I \propto 1/\gamma$ regimes.
We write Eq.~\eqref{eq:OptGam} as an approximate expression, since it only becomes exact when we take  $v \to 0$.~\footnote{Note that, even as coupling becomes weak, the current at $\gamma^\star$ in Eq.~\eqref{eq:OptGam} will not yield Landauer's result as noted in the main text after Eq.~\eqref{eq:CurrApp}, as that equation becomes exact only when the effective real--space coupling is less than the finite level spacing.} Figure~\ref{fig:AccSim} shows how the optimal estimator behaves across coupling regimes when we go to very large~$N_\qw$.  We indeed find consistency with Eq.~(\ref{eq:OptGam}) when $N_\qw$ is small and coupling is weak, observing a $\gamma^\star$ that scales as $N_\qw^{-1/2}$.   However, this  shifts to different asymptotic behavior as $N_\qw$ is increased.  The origin of this discrepancy lies in how we take limits for Eq.~(\ref{eq:CurrApp}) and subsequent results.  That is, our analytical solution is predicated on limiting to $v \rightarrow 0$ at a fixed $N_\qw$, while the procedure used in Fig.~\ref{fig:AccSim} effectively limits to $N_\qw \rightarrow \infty$ at fixed $v$.  Despite this caveat, Eq.~(\ref{eq:OptGam}) will still hold at small and moderate $N_\qw$ and thus for most practical simulations. 

Equation~\eqref{eq:OptGam} is peculiar. Folklore suggests that an appropriate $\gamma$ should be comparable to the level spacing in the reservoirs. Indeed, this assumption is used in most of the literature. For large coupling, this is correct: the linear in $\gamma$ region terminates upon reaching the Landauer plateau near a transition point at $\gamma \propto 1/N_\qw$~\cite{gruss_landauers_2016}.  The same behavior also occurs for the case discussed above.  However, owing to the anomaly, it overshoots the Landauer plateau (while both cases end at $\gamma \propto 1/N_\qw$, the numerical prefactor is different, with one being the bandwidth and the other containing the real-space system--reservoir coupling). As such, this choice can increase errors when coupling is weak. 
In other words, the relaxation defined by the level spacing $\gamma^\Hsquare = W /N_\qw$ only coincides with the optimum relaxation when $N_\qw=\frac{2\pi\omega_0^2}{v^2(1+\lambda)}$. For a smaller number of modes, $\gamma^\star$ will be weaker than $\gamma^\Hsquare$. For a larger number of modes, $\gamma^\star$ will be stronger than $\gamma^\Hsquare$. While changing the prefactor in $\gamma^\Hsquare$ can improve accuracy for finite $N_\qw$, it does not fundamentally alter this behavior. The relaxation $\gamma^\Hsquare$ will be offset from the optimum and this can result in a saturation of the relative current error at a magnitude comparable with the accuracy obtainable with tensor network simulations, see [Fig.~\ref{fig:AccSim}(c)]. 

Our observations demonstrate that, barring additional knowledge about a transport problem, one should always scan the current versus relaxation strength, as in Ref.~\cite{wojtowicz_open-system_2020}. This will indicate if anomalous behavior is present, as shown in Fig.~\ref{fig:ConfDomain} where the anomalies flank a ``domain of confidence.'' One should obtain a plateau (see Fig.~\ref{fig:AnomalousCurrents}) in this intermediate relaxation regime and extend past $k_B T/\hbar$, with $k_B$ is Boltzmann's constant and $T$ the temperature. This thermal relaxation strength marks the onset of convergence between Markovian dynamics, with an improper equilibrium, and non--Markovian dynamics, with a proper Fermi level~\cite{gruss_landauers_2016}.

The preceding results constitute an overview of anomalies for the single--site impurity, which encapsulate general implications that extend beyond this analytically tractable case. We now will discuss their physical origin and derive some approximate expressions for the current. 

\subsection{Small--$\gamma$ (virtual) anomaly}
\label{sec:Small}

\begin{figure}
\includegraphics[scale=1.00]{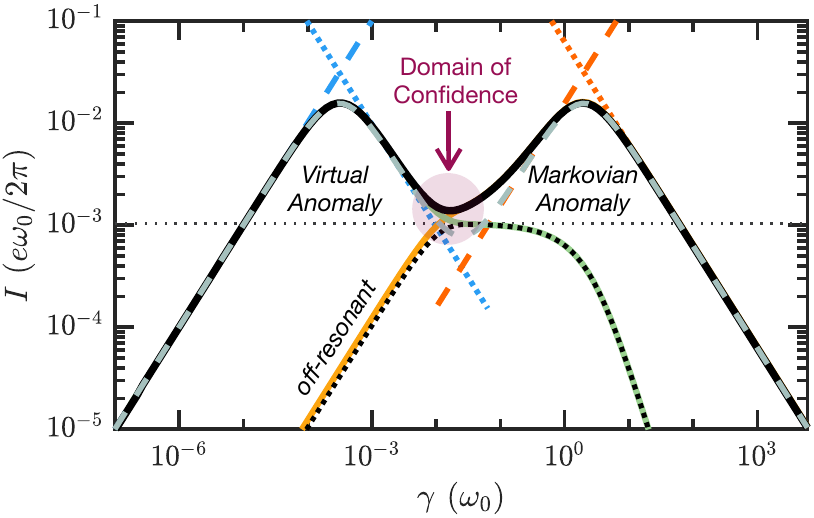}
\caption{{\bf Domain of confidence.} Turnover behavior (black solid line) generically gives rise to five regimes: (1) a region linear in $\gamma$ where the current is limited by the contact to external environment (dashed blue line); (2) a regime where virtual transitions connect resonant reservoir modes, leading to a ``virtual anomaly'' with overdamped turnover (dotted blue line); (3) a domain of confidence where calculations most closely represent the continuum limit; (4) a regime where unphysical broadening, ``the Markovian anomaly,'' creates overlap between reservoir and system states (dashed orange line); and (5) an overdamped regime where the current is inversely proportional to the relaxation (dotted orange line). The straight lines are the approximate analytic expressions for the currents in these different regimes, the dotted gray line is the Landauer limit, and the dashed gray line the ``dual'' anomalies, Eq.~\eqref{eq:CurrApp}, at weak coupling. The calculation has the same parameters as Fig.~\ref{fig:AnomalousCurrents}(b) with $N_\qw=128$ evenly spaced reservoir modes. Also shown are three other calculations: one (yellow solid line) with shifted Markovian reservoirs so that all $\ql$ modes are off resonant with $\qr$ modes, removing the virtual anomaly; a second simulation (green solid line) with non--Markovian relaxation, which contains the virtual anomaly but not the Markovian anomaly since there is a well--defined Fermi level. The third (black dashed line) has shifted, non--Markovian reservoirs, removing both anomalies. Note that the non--Markovian relaxation often displays different scaling in the large--$\gamma$ regime (here $\propto\!\! 1/\gamma^2$ since $\omega_\qs$ is outside the bias window, whereas it would give $1/\gamma$ otherwise~\cite{gruss_landauers_2016}).
\label{fig:ConfDomain}}
\end{figure}

For very small $\gamma$ we can expand Eq.~\eqref{eq:CurrApp} to obtain
\[
I \LR{\overset{\PC, \M, \WC}{\approx}} \frac{2 e \lambda}{(1+\lambda)^2} N_\qb \gamma \overset{\ES}{\approx}  \frac{2 e \mu \lambda}{W (1+\lambda)^2} N_\qw \, \gamma ,
\]
which is the weak relaxation expression when limited to linear response and weak coupling~\cite{gruss_landauers_2016,zwolak_analytic_2020,zwolak_comment_2020} [the second equality invokes the approximation in Eq.~\eqref{eq:Nmu} and the bandwidth for a 1D DOS] [see Eq.~\eqref{eq:Curr_WG_PC}]. This expression assumes that both $\vFs/\gamma \gg \gamma$ and $\vFs/\gamma \gg \left| \omega_\qs \right|$. Furthermore, since $\vFs \propto \vs/N_\qw$, the second condition will define the start of the linear regime for non--resonant tunneling at weak coupling.  This implies that $\gamma$ must be much smaller than the level spacing in the reservoirs and also much smaller than the effective coupling $\vFs/\omega_\qs$ (i.e., perturbatively from one reservoir mode to the system state). One can, however, satisfy the first inequality yet have a case where $\vFs/\gamma \ll \left| \omega_\qs \right|$.  This yields a a second expression for the current,
\[ \label{eq:SGturn}
I  \LR{\overset{\PC, \M, \WC}{\approx}} 2 e \lambda N_\qb \frac{ \left| v_F \right|^4}{\omega_\qs^2} \frac{1}{\gamma}  \overset{\ES}{\approx} \frac{8 e \mu \lambda \left| v \right|^4}{\pi^2 \omega_\qs^2 \omega_0} \frac{1}{N_\qw \, \gamma} ,
\]
corresponding to an additional $ 1/\gamma$ regime where  virtual tunneling between individual pairs of reservoir states is suppressed. That is, we now have an effective coupling of $\vFs/\omega_S$ between the system and $\ql$ (and $\lambda \vFs/\omega_S$ for $\qr$). While coherence is suppressed in this regime, similar to its large--$\gamma$ counterpart, this suppression now occurs for virtual processes. It depends on $\vf$, the reservoir size, and the system's $\omega_\qs^2$, i.e., to the perturbative coupling squared, $( \vs /\omega_\qs )^2$. Figure~\ref{fig:ConfDomain} shows where these approximations fit on the full turnover profile.

The origin of this process immediately suggests how we can remove the anomaly.  We need only shift the $\ql$ and $\qr$ modes out of alignment---and thus out of resonance---so that the virtual tunneling events are suppressed by an additional factor of the level spacing (for the equally--spaced case, the shift can be at most $W/(2 N_\qw)$). Figure~\ref{fig:ConfDomain} also shows the turnover profile with  suppressed virtual tunneling, leaving a $I \propto \gamma$ region that transitions into the large $\gamma$ anomaly directly through the intermediate, physical regime.

If $\gamma \ll k_B T/\hbar$, the non--Markovian and Markovian relaxation will have similar behavior in the weak--to--moderate $\gamma$ regime~\cite{gruss_landauers_2016}. This implies that non--Markovian relaxation will have the same virtual anomaly, and thus we do not discuss it separately. 

\subsection{Large--$\gamma$ (Markovian) anomaly}
\label{sec:Large}

For very large $\gamma$, we can expand Eq.~\eqref{eq:CurrApp} to obtain
\[
I \LR{\overset{\PC, \M, \WC}{\approx}}  \frac{4 e \lambda}{1+\lambda} N_\qb \frac{\vFs}{\gamma} \overset{\ES}{\approx} \frac{4 e \mu \lambda}{(1+\lambda)\pi \omega_0} \frac{\vs}{\gamma} ,
\]
which is the strong relaxation expression limited to linear response~\cite{gruss_landauers_2016,zwolak_analytic_2020,zwolak_comment_2020}  [again, the second equality uses the approximation in Eq.~\eqref{eq:Nmu} and the bandwidth for a 1D DOS] [see Eq.~\eqref{eq:Curr_SG_PC}]. In contrast to small--$\gamma$ approximation, this expression now assumes that $\gamma/2 \gg (1+\lambda)\vFs/\gamma$ and  $\gamma/2 \gg \left| \omega_\qs \right|$. Since $\vFs \propto \vs/N_\qw$, the second condition will define the start of the $1/\gamma$ regime for non--resonant tunneling at weak coupling. For moderate--to--strong coupling, broadening of the system mode due to system--reservoir coupling will  determine where the transition happens. For the cases when the first inequality is satisfied but the second is violated, i.e., we have $\gamma/2 \ll \left| \omega_\qs \right|$, the following relation is established:
\[ \label{eq:LGlin}
I \LR{\overset{\PC, \M, \WC}{\approx}} \frac{e \lambda}{1+\lambda} N_\qb \frac{\vFs}{\omega_\qs^2} \gamma  \overset{\ES}{\approx} \frac{e \mu \lambda}{(1+\lambda) \pi \omega_0} \frac{\vs}{\omega_\qs^2} \gamma ,
\]
corresponding to another $I \propto \gamma$ regime.  In this case, reservoir states at the Fermi level (precisely, within the bias window, which is assumed to be small) are broadened to give increased spectral weight around the system's DOS. While the mathematical origin for this is clear, the behavior is unphysical since the broadened tail of the reservoir modes should not be occupied. This region is improperly populated due to the Markovian relaxation. The expressions derived for moderate--to--strong and strong relaxation are also plotted in Fig.~\ref{fig:ConfDomain}.

Markovian and non--Markovian relaxation give different behavior in the moderate--to--strong $\gamma$ regime. If we compare Eqs.~\eqref{eq:CurrNMLR} and~\eqref{eq:CurrMLR} in linear response, we see that non--Markovian relaxation gives a current from the overlap between the reservoir DOS at the Fermi level and the (broadened) system DOS at the Fermi level, $\omega=0$.  Conversely, for Markovian relaxation,  the current is the total overlap between broadened Fermi level modes in the reservoir and the system's density of states. These Lorentzian broadened modes have long tails, which give large anomalous currents when coupling to the central region is weak.  This is compounded when broadened reservoir modes at the Fermi level reach the highly peaked system DOS. This effectively puts the occupied (unoccupied) DOS from $\ql(\qr)$ ``in resonance'' with the system level(s).  Stated in another way, the broadening due to relaxation has replaced temperature in determining the effective bias window. 

The origin of this anomaly indicates that it can be removed by using non--Markovian relaxation.  Figure~\ref{fig:ConfDomain} shows this, confirming the absence of the Markovian anomaly at large--$\gamma$. While our off--resonant example gives a large $\gamma$ turnover that scales as $I \propto 1/\gamma^2$, we would get $I \propto 1/\gamma$ for non--Markovian relaxation when the system mode is in resonance (this is an example of system--induced asymptotics).  The benefits from using non--Markovian relaxation do come at a cost:  while easy to solve for non--interacting systems, this  does not yield a time--local Lindblad master equation that is amenable to many--body simulations.  

To further explore this moderate--to--strong $\gamma$ anomaly, we can do a separate calculation at weak coupling $v$. In this case, we may ignore  reservoir--induced broadening in the system if the relaxation is moderate to large.  Such maneuvers are possible since  the explicit reservoir modes become progressively disconnected from the system due to rapid decoherence at large $\gamma$ (a fact related to the Zeno paradox).  This is not possible at weak relaxation where strong coherence remains even at weak coupling (i.e., a strong hybridization of modes). 

In this weak--coupling limit at strong relaxation, the current from Eq.~\eqref{eq:CurrMLR} becomes 
\[
I \LR{\overset{\PC, \M, \WC}{\approx}}  -\frac{e N_\qb}{\pi} \frac{\lambda}{1+\lambda} \int d\omega \, \tr \left[ \bGa_F  \Im \bG^{r}_0 \right] ,
\]
where $\bG^{r}_0$ is the bare system's Green's function []i.e., for a single--site system, $\bG^{r}_0=1/(\omega-\omega_\qs + \im \eta)$ with $\eta \to 0$ at the end of calculation].   If the system has a strong peak in the DOS around $\omega_\qs$, we get the overlap integral
\[
I \LR{\overset{\PC, \M, \WC}{\approx}} e N_\qb \frac{\lambda}{1+\lambda} \left| v_F \right|^2 \frac{\gamma}{\omega_\qs^2 + \gamma^2/4},
\]
which is equivalent to Eq.~\eqref{eq:Duality} for large $\gamma$. This derivation underscores that an unphysical spread of the occupied (and unoccupied) DOS for extended reservoir modes is causing the anomaly.  

This approach generalizes to more complex systems. For instance, we can discuss multiple non--interacting system states coupled to environments with a proportionality factor $\lambda$:
\[
I \LR{\overset{\PC, \M, \WC}{\approx}} e N_\qb \frac{\lambda}{1+\lambda} \sum_{i \in \qs} \left| v_{iF} \right|^2 \frac{\gamma}{\omega_i^2 + \gamma^2/4} .
\]
It is immediately apparent that the behavior for moderate--to--strong and strong relaxation is not markedly  different from that of a single impurity. The only difference is that  $ \left| v_F \right|^2/\omega_\qs$ is replaced by a sum over all sites in the moderate--to--strong case $ \sum_{j\in\qs}\left| v_{jF} \right|^2/\omega_j$, and furthermore that $ \left| v_F \right|^2$  is replaced with $ \sum_{j\in\qs}\left| v_{jF} \right|^2$ for the large--$\gamma$ limit.
Note that the current in these expressions scales with number of sites $N_\qs$ in the central region.  If we imagine that $\qs$ is an array  of identical sites which are decoupled from each other so that $v_{jF}=v_{F}$ and $\omega_j=\omega_{\qs}$, we will find a large-$\gamma$ anomaly that grows linearly with  $N_\qs$.

\subsection{Continuum limit}
\label{sec:Cont}

The Markovian anomaly will persist even as we approach the continuum limit, $N_\qw \to \infty$.  Conversely, the virtual anomaly will be completely suppressed, as a finite $\gamma$ is always sufficient to ``turn over'' the infinitesimal coupling from individual reservoir modes to the system. However, the level spacing will not impact moderate--to--strong relaxation, provided that it is sufficient to have anomalies separated from each other.  Since compact expressions are readily derived in the continuum limit, we can also examine behavior outside of weak coupling.  The only quantity we need is the retarded self--energy in Eq.~\eqref{eq:SECont}, which is sufficient to derive the Green's functions $\bG^r = (\omega-\omega_\qs- (1+\lambda) \bS^r)^{-1} = (\bG^a)^\dg$ and spectral densities $\bGa = -2 \Im \bS^r$. In a formal sense, this allows us to obtain expressions in  non--linear response by integrating Eq.~\eqref{eq:CurrPC}.

Nonproportional coupling may also be addressed using the more general expressions provided in Ref.~\cite{gruss_landauers_2016}.  For instance, one may assume that both reservoirs are one dimensional with a shifted DOS, as previously derived for flat--band reservoirs~\cite{gruss_landauers_2016}, to show the existence of zero--bias currents due to the Markovian anomaly. 

In linear response, we can apply Eq.~\eqref{eq:SECont} to Eq.~\eqref{eq:CurrMLR}, yielding a straightforward expression
\[ \label{eq:IMLRCont}
I \LR{\overset{\PC, \M}{\approx}} \frac{2 e \vs \mu \lambda }{\pi \omega_0 (1+\lambda)} \frac{\varUpsilon}{\omega_\qs^2+ \varUpsilon^2},
\]
where 
\[
\varUpsilon= \frac{\gamma}{2} + \frac{\vls}{2 \omega_0^2} \left( -\gamma + \sqrt{\gamma^2 + 4 \omega_0^2} \right).
\]
This result includes incoherent processes due to relaxation from implicit environments alongside coherent processes from extended environments $\ql(\qr)$. 
For convenience, we defined a modified coupling constant, $\vls = (1+\lambda) \vs$. When $\gamma$ is still weak compared to the total hybridization strength $\vs/\omega_0$, we expect a linear increase in the current versus $\gamma$.  This arises from Lorentzian broadening of the Fermi level mode in the reservoir, which increases its weight near the system DOS (this is essentially static).  The current, with its linear in $\gamma$ correction, is given by
\[ \label{eq:CurrMLRContApp}
I \LR{\overset{\PC, \M}{\approx}} I^\circ + \gamma \frac{2 e \lambda \vs \mu}{\pi (1+\lambda) \omega_0} \frac{\left(\vls - \omega_0^2 \right) \left( \vlf - \omega_0^2 \omega_\qs^2 \right)}{\left( \vlf + \omega_0^2 \omega_\qs^2 \right)^2} ,
\]
which converges to Eq.~\eqref{eq:LGlin} for weak coupling. This correction can be positive or negative depending on the alignment of the system level and the given coupling strength. For weak coupling it is a linear increase away from the Landauer limit, which will subsequently reach a maximum and decrease as $1/\gamma$.

To see the sign of the correction, we can examine the extremum in the current for $\vls > \omega_0^2$, occurring at
\[
\gamma=\frac{2 \vls -2 \omega_0^2}{\sqrt{2 \vls - \omega_0^2}}.
\]
The value of the current is given by Eq.~\eqref{eq:IMLRCont} with $\varUpsilon^2 = 2 \vls - \omega_0^2$. This is a maximum when $\omega_\qs^2 < \varUpsilon^2$ [i.e., when $\vls > (\omega_0^2+\omega_\qs^2)/2$], but a minimum when $\omega_\qs^2 > \varUpsilon^2$ [i.e., when $\vls < (\omega_0^2+\omega_\qs^2)/2$, with this condition yielding a negative sign in Eq.~\eqref{eq:CurrMLRContApp}~\footnote{We have two conditions for the minimum, $\vls > \omega_0^2$ and $\vls < (\omega_0^2+\omega_\qs^2)/2$, which together give that $\vlf < \omega_0^2 \omega_\qs^2$. The former condition also gives that $\vls-\omega_0^2>0$. Thus, the linear term is negative.}]. For increasing relaxation rate, the minimum will be followed by a subsequent maximum with current 
\[ \label{eq:Imax}
I^\top = \frac{e \lambda \vs \mu}{\pi (1+\lambda)\omega_0 \omega_\qs} 
\]
at
\[ \label{eq:gammax}
\gamma=\omega_\qs + \frac{-\omega_0^2 \omega_\qs + 2 \vls \sqrt{-2\vls + \omega_0^2 + \omega_\qs^2}}{2 \vls - \omega_0^2} .
\]
Equation~\eqref{eq:Imax} is the same as Eq.~\eqref{eq:AnomalousCurr} for weak coupling. 

When $\vls < \omega_0^2$, there will be a maximum in the current with value given by Eq.~\eqref{eq:Imax} provided that $\vls < \omega_0 \omega_\qs$. This comes at Eq.~\eqref{eq:gammax} except when $\vls = \omega_0^2/2$ where $\gamma = (4 \omega_\qs^2 - \omega_0^2)/(2\omega_\qs)$.  These results show a wide range of behavior for Markovian relaxation, even for a simple, single--site model.  Thus, when performing practical simulations, it is imperative to increase $N_\qw$ and decrease $\gamma$ in a manner that avoids this behavior. 
For any given system of interest, one should also scan the current versus $\gamma$ to ensure formation of a Landauer plateau. 

Moving beyond these features, the current will decrease as $\gamma$ becomes stronger. In effect, the Lorentzian is now so broad that the full system's state(s) are within its body, with a decay of $1/\gamma$ due to further broadening. This occurs as the full system's Green's function is getting sharper and sharper, since the rapid decoherence due to $\gamma$ effectively cuts the system off from the reservoirs.

\subsection{Landauer's regime} 
\label{sec:Land}

We can use the preceding results to constrain the required reservoir size and relaxation strength. The prior work demonstrated, in Ref.~\cite{gruss_landauers_2016}, that $\gamma \ll k_B T/\hbar$ must hold for the Markovian approximation (which has an ill-defined Fermi level) to converge to the non--Markovian relaxation (which has a well-defined Fermi level). This is a useful condition since neither the proof nor the expression rely on the system architecture, the reservoir band structure, or any other details, just the nature of the relaxation. Here, we give more precise conditions for our reference impurity problem, which are generally helpful for understanding the extended reservoir approach. 

We begin from the large--$\gamma$ side. To fully remove the Markovian anomaly, we require a $\gamma$  small enough that the linear component of Eq.~\eqref{eq:CurrMLRContApp} is negligible. Taking the ratio of the linear component to~$I^\circ$ to obtain an upper bound on $\gamma$ yields
\[ \label{eq:UB}
\gamma \ll \frac{\vls \omega_0 \left( \vlf + \omega_0^2 \omega_\qs^2 \right)}{\left| \vls - \omega_0^2 \right| \left| \vlf - \omega_0^2 \omega_\qs^2 \right|} \xrightarrow[]{v \to 0} \frac{\vls}{\omega_0} .
\]
Once again, the linear correction can be negative due to different behavior in different parameter regimes. 
Similarly, we can find where the virtual anomaly turnover, Eq.~\eqref{eq:SGturn}, equals $I^\circ$ to provide a lower bound on $\gamma$. In the weak--coupling limit, 
\[ \label{eq:LB}
\gamma \gg \frac{4 \omega_0}{\pi N_\qw} \overset{\ES}{=} \frac{\Delta_0}{\pi} ,
\]
where $\Delta_0$ is the level spacing. 

There is no contradiction between this result and the square--root result. Putting the two inequalities, Eqs.~\eqref{eq:UB} and~\eqref{eq:LB}, together and maximally satisfying the constraints will yield the same $N_\qw$ dependence of the optimal estimator in Eq.~\eqref{eq:OptGam}.   As we discussed earlier, Fig.~\ref{fig:AccSim} shows that the optimal estimator behaves as $1/\sqrt{N_\qw}$ for strong scattering (weak coupling) and shifts to $1/N_\qw$ for weak scattering.

\subsection{Many body $\qs$} 
\label{sec:Manybody}
\begin{figure}[th!]
\includegraphics[width=\columnwidth]{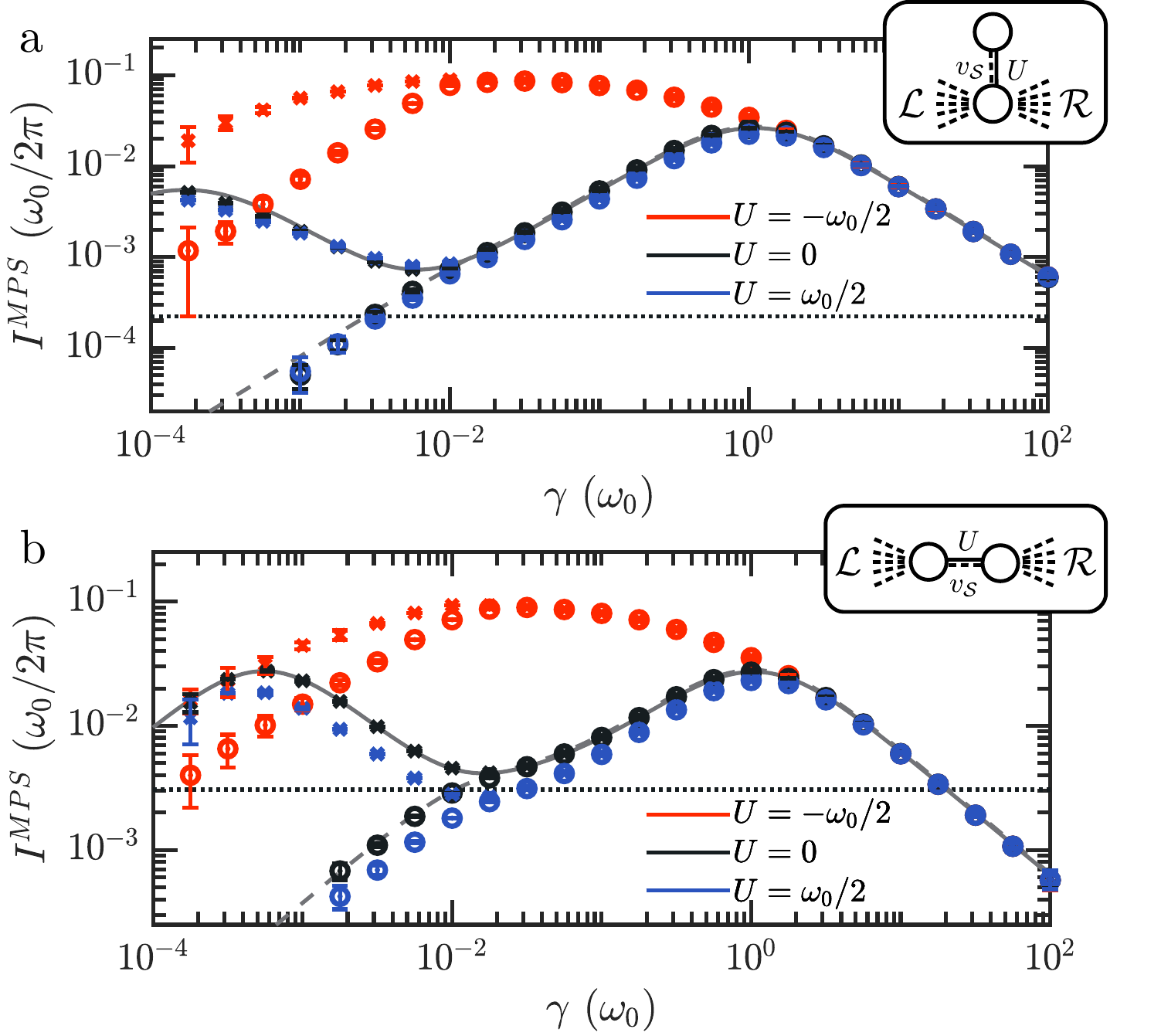}
\caption{{\bf Anomalies in interacting systems.} Current for the spinless, interacting two--site system in Eq.~\eqref{eq:H2sites}, where the insets indicate how the system is connected to the reservoirs. 
The anomalous regions are apparent and we refer to the main text for further discussion.
 Profiles with resonant and off--resonant modes are marked with open circles and crosses, respectively, with the latter always removing the virtual anomaly.
The setup follows Sec.~\ref{sec:Model} with $N_\qw=128$ evenly spaced reservoir modes. Our reservoir--impurity coupling is weak, $v=\omega_0/8$, and the inter--impurity coupling is $v_S=(1+\sqrt 2)\omega_0/4$. We apply a symmetric bias $\mu = \omega_0/2$  between $\ql$ and $\qr$ and assume a temperature $k_B T = \hbar \omega_0 / 40$.  The density--density interaction strength is $U=-\omega_0/2$ (red), $U=0$ (black), and $U=\omega_0/2$ (blue). The black lines indicate the reference (exact) solution for the non--interacting case with the horizontal line marking the Landauer limit in this case.
The error bars at $\pm \sigma$ show an estimate of the MPS convergence, where $\sigma^2=\sigma_1^2+\sigma_2^2$. Here, $\sigma_1$ is a standard deviation reflecting fluctuations of the current in a time--window of $\Delta t = 50/\omega_0$ at long times after reaching the steady--state. The current $I$ is defined as $I = \sum_i I_i/n_I$ [with $i$ iterating over $n_I=2$ possible interfaces for (a) and 3 for (b)] and $\sigma_2^2 = \sum_i |I_i-I|^2/n_I$ quantifies the mismatch of currents at different interfaces. The errors are associated with the finite MPS bond dimension, where we fix $D_{\rm max} = 256$ (or $352$ for low--$\gamma$ slopes) and truncate the MPS Schmidt values below $10^{-6}$ (whichever gives a smaller bond dimension at a given cut).
\label{fig:CDint}}
\label{fig:mps}
\end{figure}

The considerations discussed in this work carry over to more complicated non--interacting systems, as well as to interacting many--body systems. Here, we present data for the two--site systems of Eq.~\eqref{eq:H2sites} and Sec.~\ref{sec:Model}  in a weak--coupling regime.  In doing so, we compare limits with and without a  density--density interaction $U$. These data are for (i) a proportional coupling case, where both reservoirs are coupled to the same system site, and none coupled to the other system site [Fig.~\ref{fig:CDint}(a)] and (ii) for a non--proportional coupling case, where each reservoir is connected to a different site of the system  [Fig.~\ref{fig:CDint}(b)]. The reservoirs themselves are identical in both cases.

To address many--body interactions, we employ the extended--reservoir {\rm tensor--network} approach of Ref.~\cite{wojtowicz_open-system_2020}.  In this case, the density matrix of the $\ql\qs\qr$ system is vectorized and expanded as a matrix product state (MPS) using local operator bases. These are formed by operators appearing in the Lindbladian from Eq.~\eqref{eq:fullMaster} when using an energy representation~\cite{zwolak_mixed-state_2004}.  This arrangement is configured in the {\em mixed basis}, where we order modes in a manner that localizes correlations to the bias window and thus minimizes entanglement along the chain~\cite{rams_breaking_2020}.  It should be noted that the Lindbladian does not mix different particle sectors. We enforce the resulting block structure of the density matrix by forming MPS from U(1)--symmetric (particle--number--preserving) tensors~\cite{singh_tensor_2010,*singh_tensor_2011}, speeding up and stabilizing the simulations. 

The superoperator $\mathbb{L}$ encoding the action of the Lindbladian on the vectorized density matrix is represented as a matrix product operator.  This allows us to simulate Eq.~\eqref{eq:fullMaster} efficiently using the time--dependent variational principle for MPS~\cite{haegeman_time-dependent_2011,haegeman_unifying_2016}, despite the complicated long--range coupling structure in that setup (and Jordan--Wigner strings). We note that standard approaches to target the steady state directly, e.g., variationally minimizing $\mathbb{L}^\dagger \mathbb{L} $ with the help of the density--matrix renormalization group algorithm, prove to be unstable.  This behavior is due to a gap that rapidly vanishes with decreasing $\gamma$ and nontrivial entanglement that persists in the optimized basis. Nevertheless, simulating the time--evolution is a viable strategy to reach the desired steady--state.  We illustrate this in Fig.~\ref{fig:CDint} where, for $U=0$, we compare the exact values of the current (black lines) with the MPS results (black symbols). These data overlap tightly across several orders of magnitude in $\gamma$, and only in the limit of very low $\gamma$ do errors from the MPS become noticeable. A similar comparison cannot be made  for the interacting case, as exact results are inaccessible.  Thus, we must rely solely on our MPS simulations.

The results for a repulsive interaction ($U=\omega_0/2$, blue points in Fig.~\ref{fig:CDint}) closely mimic the turnover structure of the non--interacting case: both anomalies are clearly visible for unshifted modes (crosses) and only the Markovian anomaly remains when modes in $\ql(\qr)$ are brought out of alignment (circles).  Conversely, when the interaction is attractive ($U=-\omega_0/2$, red points in Fig.~\ref{fig:CDint}), the system is close to resonance is enhanced in both setups. There is now an effective impurity mode that is pushed into the bias window, hiding the two anomalies under a larger intrinsic current.  We nonetheless retain a signature of the low--$\gamma$ anomaly if the mode energies are taken out of alignment, as shown by the departure of the red crosses and red circles around $\gamma \approx \omega_0/100$.

A comparison of transport profiles when reservoirs modes are on-- and off resonant can be used to estimate an optimal $\gamma^\star$.  In particular, the Landauer regime may be found by seeking the intersection between turnover profiles with on--resonant and off--resonant modes, guiding convergence to the Landauer limit for finite--$N_\qw$ simulations.  We explore this possibility in detail in Ref. \cite{elenewski_performance_2021} while quantifying the impact of different reservoir discretization on simulation efficiency.

\section{Conclusions}
\label{sec:Conc}
The conclusions herein, together with prior results on quantum electron~\cite{gruss_landauers_2016,elenewski_communication_2017,gruss_communication_2017,zwolak_comment_2020,zwolak_analytic_2020} and classical thermal~\cite{velizhanin_crossover_2015,chien_thermal_2017} transport, provide a global perspective on the factors for  accurate extended reservoir simulations:

(i) The Markovian relaxation strength must be weaker than the thermal relaxation $k_B T/\hbar$, at least around the Fermi level, to approximate proper reservoir equilibria~\cite{gruss_landauers_2016}. This requirement can be relaxed on occasion, since transport properties do not necessarily change when passing below a given effective temperature scale. Thus, accurate conductance values will still result if $\gamma < k_B T^\star /\hbar$, with $T^\star$ the lowest of these temperature scales. This behavior is observed for common models (see, e.g., Ref.~\cite{wojtowicz_open-system_2020} where below about $k_B T^\star \approx \omega_0/10$ the conductance does not change). If there are no features, whether in the density of states, the transmission function, or due to many-body interactions (e.g., the Kondo temperature \cite{hewson_kondo_1993}), then there is no reason to have an excessively small $\gamma$. This is heuristic and should be employed with care, especially when considering other observables such as noise. 

It should be noted that there is a fundamental pathology of the Markovian Lindblad master equation, Ref.~\cite{tupkary_fundamental_2021}, which is due to to improper equilibrium (i.e., the Markovian equation relaxes the modes to the occupation of the equilibrium state for isolated reservoirs). However, it can be rigorously proven that with appropriate choice of parameters an approach based on the Lindblad equation limits to the correct results. The approach based on an extended reservoirs and a Redfield master equation is a promising alternative which might potentially help convergence to the thermodynamic limit for some cases. 

(ii) Other energy scales appear within the Markovian extended reservoir framework. In particular, we have shown that the reservoir spacing, $\Delta$, and the coupling, $\vs/\omega_0$ (or its more complicated version), set  important energy scales for simulation. These bound the domain of confidence for $\gamma$ from below and above, respectively. The former suppresses virtual transitions and thus helps to identify an upper bound for the virtual anomaly. The latter ensures that broadening is under control and that the simulation result lies prior to the Markovian anomaly. 

(iii) Certain features, such as the band structure and gap states~\cite{gruss_communication_2017,chien_thermal_2017}, can collectively give rise to a range of behavior. Many--body interactions may {\em sometimes} help in this regard (though they may also be detrimental) by smearing sharp features in the DOS, particularly those that support interference or effects that give rise to anomalous currents. 

The only generally effective approach for extended reservoir transport simulations  is to scan physical observables versus relaxation strength and reservoir size.  This  consideration also holds for other applications, such as non--equilibrium thermodynamics.  Weak--coupling cases are particularly troublesome due to the anomalies studied here and the complexity resulting from five distinct transport regimes.  Other methods, such as perturbative treatments of the system--reservoir coupling, can address parameter regions where extended reservoirs may be numerically difficult to apply.~\footnote{If one desires to simulate cases where contact is weak, or where it fluctuates between weak and strong regimes (such as Floquet states that are modulated by a harmonic function), then one needs to resolve this issue and have a verified and validated simulation approach throughout the range of parameters (e.g., large enough reservoir size, small enough gamma that one remains on the plateau).  Similar concerns arise when discussing electronic sensing and junction configurations~\cite{zwolak_electronic_2005,lagerqvist_fast_2006,lagerqvist_comment_2007,lagerqvist_influence_2007,zwolak_colloquium_2008,tsutsui_identifying_2010,huang_identifying_2010,chang_electronic_2010,ohshiro_single-molecule_2012}.} It remains to be seen if these factors influence alternative implementations with intermode relaxation in the reservoirs.

Our results provide a comprehensive perspective for extended reservoir simulations, while refining the Kramers' turnover picture for open transport simulations.  We likewise establish a domain of confidence where Markovian relaxation can provide accurate transport profiles.  This will advance the practical use of these methods and help usher in the era of ERAs.

J.~E.~E. acknowledges support under the Cooperative Research Agreement between the University of Maryland and the National Institute for Standards and Technology Physical Measurement Laboratory, Award 70NANB14H209, through the University of Maryland. We acknowledge support by the National Science Center (NCN), Poland under Projects No. 2016/23/B/ST3/00839 (G.~W.) and No. 2020/38/E/ST3/00150 (M.~M.~R.).

\end{document}